\documentclass[11pt]{article}
\usepackage{graphicx,amssymb}

\begin{document}
  
  \title{Mass Functions of Supermassive Black Holes Across Cosmic Time}

  \author{Brandon C. Kelly\thanks{Hubble Fellow,
      bckelly@cfa.harvard.edu. Current address: Dept. of Physics,
      Broida Hall, University of California, Santa Barbara, CA 93106},
    Harvard-Smithsonian Center for Astrophysics, \\
    60 Garden St, Cambridge, MA 02138 USA
    \and Andrea Merloni
    Max Planck-Institut f\"{u}r Extraterrestrische \\
    Physik, Giessenbachstrasse 1, D-85748 Garching, German}

  \maketitle

  \begin{abstract}
    The black hole mass function of supermassive black holes describes
    the evolution of the distribution of black 
    hole mass. It is one of the primary empirical tools available for
    mapping the growth of supermassive black holes and for constraining
    theoretical models of their evolution. In this review we discuss
    methods for estimating the black hole mass function, including
    their advantages and disadvantages. We also review the
    results of using these methods for estimating the mass function of
    both active and inactive black holes. In addition, we review
    current theoretical models for the growth of supermassive black
    holes that predict the black hole mass function. We conclude with a
    discussion of directions for future research which will lead to
    improvement in both empirical and theoretical determinations of
    the mass function of supermassive black holes.
  \end{abstract}
  
  \section{INTRODUCTION}
  
  \label{s-intro}
  
  Understanding how and when supermassive black holes (SMBHs) grow is
currently of central importance in extragalactic astronomy. A
significant amount of empirical work has established correlations
between SMBH mass and host galaxy spheroidal properties, such as
luminosity \cite{korm95,mclure01,mclure02}, stellar velocity
dispersion (the $M_{BH}$--$\sigma_*$ relationship, e.g., \cite{gebh00a,
merr01, trem02}), concentration or Sersic index
\cite{graham01,graham07}, bulge mass
\cite{mag98,marc03,haring04}, and binding energy
\cite{aller07,fundplane}. These scaling relationships imply
that the evolution of spheroidal galaxies and the growth of SMBHs are
intricately tied together. The currently favored mechanism for linking
the growth of SMBHs and their hosts is black hole feedback, whereby
black holes grow by accreting gas in so-called ``active'' phases, possibly fueled by a major merger
of two gas-rich galaxies, until feedback energy from the SMBH expels
gas and shuts off the accretion process
\cite{silk98,fabian99,begel05,murray05}. Alternatively,
it has been suggested that the origin of the scaling relationships
does not necessarily require SMBH feedback, but emerges from the
stochastic nature of the hierarchical assembly of black hole and stellar mass through
galaxy mergers. \cite{peng07,jahnke11}.

Feedback-driven `self-regulated' growth of black holes has been able
to reproduce the local $M_{BH}$--$\sigma_*$ relationship in smoothed
particle hydrodynamics simulations
\cite{dimatteo05,spring05,johan09}. Moreover, AGN feedback has also
been invoked as a means of quenching the growth of the most massive
galaxies \cite{bower06,croton06}. There have been numerous models
linking SMBH growth, the quasar phase, and galaxy evolution
\cite{haehn98,kauff00,haehn00,wyithe03,vol03,catt05,dimatteo08,
somer08,booth09}. While feedback is likely important for regulating
the growth of SMBHs and galaxies, the fueling mechanisms that
contribute to growing the SMBH are likely diverse. Major-mergers of
gas-rich galaxies may fuel quasars at high redshift, and grow the most
massive SMBHs. However, major-mergers alone do not appear to be
sufficient to reproduce the number of X-ray faint AGN
\cite{marulli07}, and accretion of ambient gas via internal galactic
processes \cite{ciotti01,stochacc}, may fuel these fainter, lower
$M_{BH}$ AGN at lower $z$. This is supported by the fact that many AGN
are observed to live in late-type galaxies out to $z \approx 1$
\cite{guyon06,gabor09}, and the X-ray luminosity function of AGN
hosted by late-type galaxies suggests that fueling by minor
interactions or internal instabilities represents a non-negligible
contribution to the accretion history of the Universe \cite{georg09}.

The black hole mass function (BHMF) provides a complete census of the
mass of SMBHs and their evolution. Because of this, the BHMF is one of
the primary empirical tools available for investigating the growth of
SMBHs, and for constraining theoretical models for the growth of the
SMBH population. Because SMBHs and galaxies are thought to be linked
in their evolution, the BHMF provides insight into the fueling
mechanisms that dominate black hole growth, and therefore into the
role of feedback in the evolution of the host galaxy. The BHMF is also an important tool in planning future surveys, as it
provides an estimate of the distribution of SMBH mass expected for the
survey. This in turn is important because mass is a fundamental
quantity of the black hole, and therefore is an important
observational quantity for empirical studies of black hole accretion
physics
\cite{merloni03,falcke04,mchardy06,davis11,sob11,kelly11}. Of
course, further improvement to our understanding of black hole
accretion physics will further improve our modeling and understanding
of black hole accretion and feedback, which in turn will improve our understanding
of black hole-galaxy coevolution. Therefore, the BHMF is an
important empirical quantity for SMBH studies. 

In this review we discuss the current status of BHMF estimation and
theoretical modeling. In \S~\ref{s-est} we discuss the non-trivial
task of estimating the BHMF. In \S~\ref{s-local} we discuss current
estimates of the local SMBH BHMF. In \S~\ref{s-continuity} we discuss
BHMF estimates derived by combining the local BHMF with the AGN
luminosity function via a continuity equation. In \S~\ref{s-agn} we
discuss BHMFs estimated for AGN only. In \S~\ref{s-theory} we review
theoretical models for SMBH growth that predict the SMBH
BHMF. Finally, in \S~\ref{s-future} we discuss directions for future
improvements to the empirical and theoretical studies of the BHMF. We
note that unlike, say, the luminosity function, the division between
`observational' and `theoretical' studies is not as clear for the
BHMF, as some amount of modeling is necessary in order to estimate the
BHMF from strictly observational quantities. We have attempted to
divide the studies according to whether the BHMF is constrained
empirically, as in, say, a formal statistical fitting procedure, or if
it is predicted from a theoretical model for SMBH growth. In reality, the line
between theoretical and empirical studies is blurry, and some
procedures which we have considered to be empirical may be thought of
as theoretical. 
  
  \section{Estimating the Black Hole Mass Function}
  
  \label{s-est} 
  
  The black hole mass function, denoted as $\phi(M_{BH}, z) dM_{BH}$,
  is the number of sources per comoving volume $V(z)$ with black hole
  masses in the range $M_{BH}, M_{BH} + dM_{BH}$. The black hole mass
  function is related to the joint probability distribution of $M_{BH}$ and $z$, $p(M_{BH},z)$, as
  \begin{equation}
    \phi(M_{BH},z) = N \left(\frac{dV}{dz}\right)^{-1} p(M_{BH},z).
    \label{eq-phiconvert}
  \end{equation}
  The normalization of the BHMF is $N$, the total number of SMBHs in
  the observable universe, and is
  given by the integral of $\phi$ over $M_{BH}$ and $V(z)$. 

  \subsection{Complications with Estimating Black Hole Mass Functions}

  \label{s-complications}

  Similar to luminosity function estimation, the BHMF may be estimated
  from astronomical surveys. However, while there are many
  well-established methods for estimating luminosity functions, there
  are two complications that make BHMF estimation a more difficult
  problem \cite{kvf09}. The first is the issue of
  incompleteness. Surveys are typically constructed by finding the set
  of objects of interest containing a SMBH that satisfy a flux criteria,
  e.g., all objects brighter than some flux limit. Surveys are not
  constructed by selecting on mass. Because there is a distribution of
  luminosities at a given SMBH mass, whether it be the luminosity of the
  host galaxy or of the AGN, some SMBHs will scatter above the flux limit and
  some below. This creates a selection function which is less sensitive
  to $M_{BH}$, and it is possible that a survey may be incomplete in all
  mass bins.

  The second complication is the large uncertainty in SMBH mass
  among mass estimators. Currently it is not possible to obtain reliable mass
  estimators for large numbers of SMBHs through dynamical and modeling of the stellar or gaseous components, and thus scaling relationships are
  employed. Masses may be estimated using scaling relationships
  between $M_{BH}$ and the properties of the host galaxy bulge or the luminosity and the width of the broad
  emission lines for AGN \cite{wandel99,vest06}. It has also recently been
  suggested that the X-ray variability properties of AGN may also
  provide another scaling relationship for estimating $M_{BH}$
  \cite{nik04,zhou10,kelly11}, but 
  further work is needed for developing this. While these scaling
  relationships enable one to estimate $M_{BH}$ for large numbers of
  SMBHs, they also contain a significant intrinsic statistical
  scatter. G{\"u}ltekin, K., 
    et al. \cite{gult09b} find that for early type galaxies there
  is an intrinsic scatter in $M_{BH}$ of $0.31 \pm 0.06$ dex and $0.38
  \pm 0.09$ dex at fixed host galaxy bulge dispersion and luminosity,
  respectively; the amplitude of the scatter is larger for late type
  galaxies. For AGN with broad emission lines, Vestergaard \& Peterson \cite{vest06} estimate
  the scatter in $M_{BH}$ at fixed luminosity and line width to be
  $\sim 0.4$ dex, depending on which emission line is used.

\begin{figure}[t]
  \includegraphics[scale=0.5,angle=90]{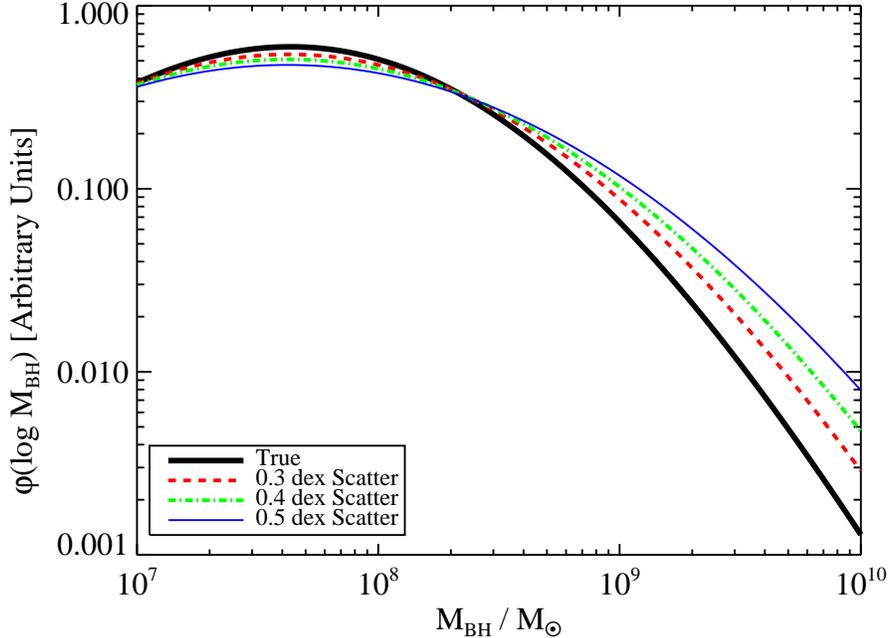}
  \caption{Illustration of the bias in the estimated BHMF derived from
    mass estimates. Shown is the true mass function (thick solid black
    line) for a simluated sample, and the mass function derived from
    the mass estimates when the statistical error in the mass
    estimates is 0.3 dex (red dashed line), 0.4 dex (green dot-dashed
    line), and 0.5 dex (solid blue thin line). The mass function
    estimated from the mass estimates is biased, especially at the
    high $M_{BH}$ end and for large statistical error.}
  \label{f-bhmf_illust}
\end{figure}

  The statistical uncertainty in the mass estimates can have a
significant effect on the inferred BHMF. The distribution of the mass
estimates is the convolution of the intrinsic BHMF with the error
distribution in the mass estimates. In general, it is typically
assumed that the error in the mass estimates is independent of the
actual value of $M_{BH}$. This is not the case for $M_{BH}$ estimated
through dynamical modeling, however independence between $M_{BH}$ and
its error is likely to be a good
approximation for $M_{BH}$ estimated using scaling
relationships. Because scaling relationships are the only feasible
manner in which to estimate $M_{BH}$ for a large sample of SMBHs,
which is necessary for any estimate of the BHMF, we will assume that
$M_{BH}$ and its error are independent. Under the assumption of
independence between the estimated $M_{BH}$ and its error, the BHMF that would
be inferred directly from the distribution of the mass estimates is
broader than the intrinsic BHMF, and is thus biased. Figure
\ref{f-bhmf_illust} illustrates this effect, where an intrinsic mass
function is compared with the distribution of an unbiased mass
estimator having a statistical uncertainty of 0.3, 0.4, and 0.5 dex,
respectively. As can be seen, the distribution of mass estimates is
significantly different than the intrinsic mass function. In
particular, the distribution of the mass estimates falls off more
slowly with increasing $M_{BH}$, and overpredicts the number of SMBHs
at the high $M_{BH}$ end of the mass function. The bias is worse when
the dispersion in the scatter in the mass estimates becomes larger.

  \subsection{Methodology for Estimating the Black Hole
    Mass Function}

  \label{s-method}

  In order to estimate the SMBH mass function in an unbiased manner,
  it is necessary to match the mass function with the observed
  distribution of the mass estimates and any additional observational
  quantities that the selection function\footnote{The selection function is the probability of including a source in one's sample as a function of its measured quantities.} depends on. The basic idea is
  to start with an assumed mass function. Then, calculate the
  distribution of mass estimates implied by this mass function. In
  addition, calculate the distribution of observational quantities
  that one's sample is selected on, say, flux, that is implied by the
  assumed mass function. This step allows one to correct for
  incompleteness, but requires an additional assumption about how to
  relate the mass function to the quantity that one's sample is
  selected on. Finally, impose the selection function for the sample,
  and compare the predicted observed distributions of mass estimates
  and any other observables (e.g., flux) with the actual
  distributions. If they are not consistent, then the data rule out the assumed mass
  function and relationship between $M_{BH}$ and the observable quantities. 

  We can make the above procedure more quantitative by deriving the
  likelihood function for the SMBH mass function. Kelly, Vestergaard, \& Fan \cite{kvf09}
  derived the likelihood function for the mass function when using
  masses estimated from AGN broad emission lines. They used this
  likelihood function for developing a Bayesian approach to estimating
  the SMBH mass function. Although their method was limited to broad
  line mass estimates, it is straightforward to generalize their
  formalism for any generic mass estimator. Denote the black hole mass
  estimate as $\hat{M}_{BH}$. In addition, denote as $X$ the set of
  observables that one uses to select one's sample. In the majority of
  cases this will be flux at one or more wavelengths. Then, the
  likelihood function for the BHMF based on a sample of $n$ SMBHs is
  \begin{eqnarray}
    \lefteqn{p(\hat{\bf M}_{BH},{\bf X},{\bf z}|\theta,\psi,N) \propto} \nonumber \\
    & & C^N_n \left[ s(\theta,\psi) \right]^{N-n}
    \prod_{i=1}^n \int_{0}^{\infty} p(\hat{M}_{BH,i}|M_{BH,i}, z_i,
    X_i) p(X_i|M_{BH,i},z_i,\psi) p(M_{BH,i},z_i|\theta)\ dM_{BH,i}, 
    \label{eq-likhood}
  \end{eqnarray}
  where the BHMF is related to $N$ and the probability distribution of
  $M_{BH}$ and $z$ via Equation (\ref{eq-phiconvert}). Here, $C^N_n$
  is the binomial coefficient, $\theta$ denotes the parameters for the
  BHMF, $\psi$ denotes the parameters for the distribution in $X$ at
  fixed $M_{BH}$ and $z$, and $s(\theta,\psi)$ is
  the probability of including a SMBH in one's sample as a function $\theta$ and
  $\psi$. Here, we have assumed that the distribution in the mass
  estimates at fixed $M_{BH},z,$ and $X, p(\hat{M}_{BH}|M_{BH},z,X)$,
  is known, although one could include additional free parameters for
  this as well. The binomial coefficient arises from the fact that the
  number of objects included in one's survey follows a binomial
  distribution\footnote{Often in the luminosity function
    literature the likelihood is assumed to be a Poisson
    distribuiton. A poisson distribution is an approximation to the
    binomial distribution when $N \rightarrow \infty$ and
    $s(\theta,\psi) \rightarrow 0$, so Equation (\ref{eq-likhood})
    converges to the Poisson distribution when $n \ll N$. See
    \cite{kfv08} for further details.} with $N$ 'trials' and probability of success
  $s(\theta,\psi)$. The probability of including a SMBH in one's survey as a function of
  the BHMF, $s(\theta,\psi)$, is calculated from the survey selection function
  $s(X,z)$ as 
  \begin{equation}
    s(\theta,\psi) = \int_{X_{min}}^{X_{max}} s(X,z) \left[ \int_{0}^{\infty}
      \int_{0}^{\infty} p(X|M_{BH},z,\psi) p(M_{BH},z|\theta)\
      dM_{BH}\ dz \right]\ dX. 
    \label{eq-selfunc}
  \end{equation}
  It is up to the researcher to choose the particular parameteric form
  for the SMBH mass function, the distribution in the mass estimates
  at fixed $M_{BH}, z,$ and $X$, and the distribution of the
  observable that the sample is selected on (e.g., flux) at fixed $M_{BH}$ and
  $z$. Typical choices are log-normal distributions, Schechter functions, and mixtures of log-normal distributions. Once one has done this, one can use Equation (\ref{eq-likhood})
  to compute a maximum-likelihood estimate for the BHMF, or perform
  Bayesian inference. 

  An alternative form of estimating the BHMF can be used when the mass
  estimates are derived from an observational quantity, $Y$, and the
  intrinsic distribution of $Y$ is known. This is commonly used to
  estimate the local mass function using host-galaxy scaling
  relationships \cite{marc04}. In this case, the mass
  function is 
\begin{equation}
  \phi(M_{BH}) = \int_{Y_{min}}^{Y_{max}} p(M_{BH}|Y) \phi(Y)\ dY,
  \label{eq-mfunc_alt}
\end{equation}
where $\phi(Y)$ is the comoving number density of SMBHs as a function
of the quantity $Y$. When both $p(M_{BH}|Y)$ and $\phi(Y)$ are known
then the BHMF follows directly from Equation (\ref{eq-mfunc_alt}). As
an example, if the mass function is derived from the scaling between
$M_{BH}$ and host galaxy spheroidal luminosity, $L_{sph}$, then $Y =
L_{sph}$, $p(M_{BH}|Y)$ is the $M_{BH}$--$L_{sph}$ relationship, and
$\phi(Y)$ is the luminosity function of stellar bulges hosting
SMBHs. As with BHMFs determined from a mass estimator, improper
treatment of the intrinsic scatter in $M_{BH}$ at fixed $Y$ will lead
to a biased estimate of the BHMF. However, when calculating the BHMF
from Equation (\ref{eq-mfunc_alt}), ignoring the intrinsic scatter
results in an estimated BHMF that is too narrow,  underpredicting the
number of SMBHs at the high mass end of $\phi(M_{BH})$.  This is opposite to the case when one estimates the BHMF directly from mass estimates.

  \section{Black Hole Mass Functions Derived from Host Galaxy
    Scaling Relationships}

  \label{s-local}

  The observed scaling between $M_{BH}$ and the properties of the SMBH
  host galaxy bulge have motivated several groups to estimate the
  local BHMF
  \cite{sal99,yu02,aller02,marc04,shank04,lauer07,graham07,tundo07,yu08,merloni08,shank09,vika09},
  with decreasing statistical uncertainties. These estimates of the
  local BHMF have formed the basis for many studies which have
  attempted to map black hole growth by comparing with the AGN
  luminosity function; this is further discussed in
  \S~\ref{s-continuity}. Typically, the local BHMF is estimated using
  the local $M_{BH}$--$\sigma_*$ relationship or the local
  $M_{BH}$--$L_{sph}$ relationship, combined with the local number
  density of galaxies as a function of stellar velocity dispersion or
  bulge luminosity. 

  The scaling relationships between $M_{BH}$ and host galaxy
properties are only determined for the local universe, and thus most
authors have limited their determination of the BHMF based on them to
the local BHMF. An exception is Tamura, Ohta, \& Ueda \cite{tam06},
who estimated the BHMF out to $z \approx 1$ assuming that evolution in
the $M_{BH}$--$L_{sph}$ relation is driven only by passive evolution
in $L_{sph}$.  Evolution in the scaling relationships is currently an
area of intense study, with most groups finding evidence that the
normalization of the scaling relationships increases towards higher
$z$ \cite{treu04,peng06,treu07,woo08,merloni10,bennert10}, at least for active SMBHs. However,
there are still concerns regarding potential biases due to selection
effects \cite{lauer_bias}, but see Treu et al. \cite{treu04} and
Bennert et al. \cite{bennert10} for procedures aimed at modeling and
correcting for selection. There may also be biases due to
extrapolating the AGN mass estimates derived from the broad emission
lines to luminous quasars at high $z$ \cite{shen_kelly10}. As such,
the uncertainties on the quantitative form of the evolution in the
scaling relationships and their scatter are currently large, limiting
their use for determining the BHMF outside of the local universe.

\begin{figure}[t]
  \includegraphics[scale=0.5,angle=90]{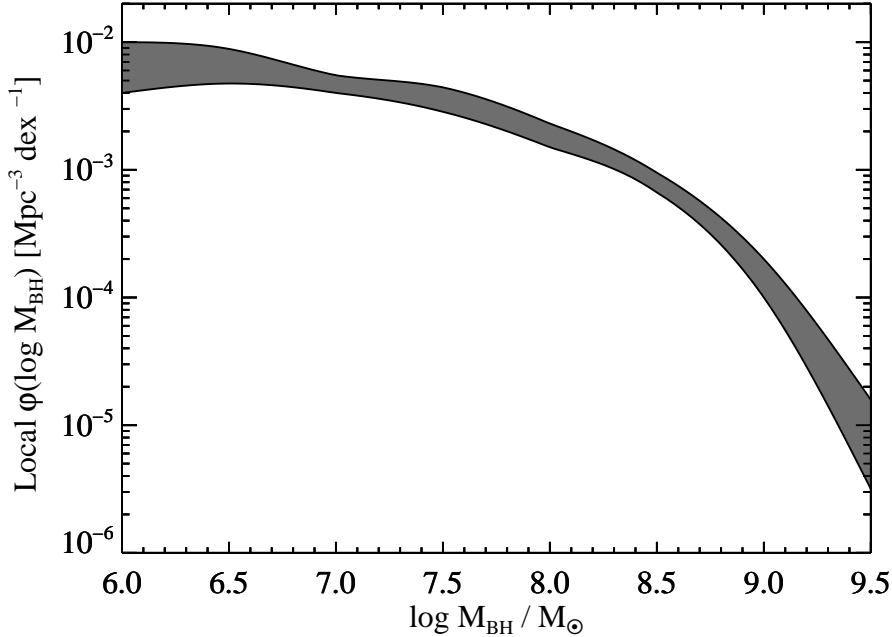}
  \caption{Local BHMF. The shaded region defines the spread in estimates obtained using the $M_{BH}$--$\sigma_*$, $M_{BH}$--$L_{sph}$, and $M_{BH}$--$M_{\rm star}$ relationships, as compiled by Shankar et al. \cite{shank09}. Based on this estimate the local
    universe is dominated by SMBHs with $M_{BH} < 10^7 M_{\odot}$.}
  \label{f-local_bhmf}
\end{figure}

When the $M_{BH}$--$\sigma_*$ relationship is used to estimate the
local BHMF, it is common to use the velocity dispersion distribution
derived from the SDSS by Sheth et al. \cite{sheth03}, with an additional component
representing the brightest cluster galaxies
\cite{lauer07}. Sheth et al. \cite{sheth03} estimate the velocity dispersion
distribution for late type galaxies by using the Tully-Fisher relation
to convert the luminosity function of late type galaxies to a circular
velocity distribution, and then set $\sigma = v_c / \sqrt{2}$. When
the $M_{BH}$--$L_{sph}$ relation is used it is typical to estimate the
distribution of $L_{sph}$ seperately for early and late type galaxies
by converting their respective luminosity functions to spheroidal
luminosity functions using an assumed ratio of bulge luminosity to
total luminosity. From this it has been inferred that the local BHMF
is dominated by early type galaxies at $M_{BH} \gtrsim 4 \times 10^7
M_{\odot}$ \cite{yu08}. Shankar, Weinberg, \& Miralda-Escud{\'e} \cite{shank09} present a compilation of
recently determined local BHMFs based on a variety of methods, scaling
relations used, and data sets used. In Figure \ref{f-local_bhmf} we
show the range of local BHMFs estimated from the $M_{BH}$--$\sigma_*$, $M_{BH}$--$L_{sph}$, and $M_{BH}$--$M_{\rm star}$ relationships, as presented in Shankar et al. \cite{shank09}. In
general, estimates of the local 
mass density of SMBHs span the range $\rho_{BH} = (3.2$--$5.4) \times
10^5 M_{\odot} {\rm\ Mpc}^{-3}$ for $h = 0.7$ \cite{shank09}. 

While the procedure for estimating the local BHMF is, in theory,
straight-forward, a number of significant systematics remain. First,
there is the observational difficulty that most BHMFs derived from
the $M_{BH}$--$\sigma_*$ relationship are based on SDSS
spectra. Unfortunately, the SDSS velocity dispersions are based on a
fixed aperture, and thus the size of the aperture relative to the bulge varies with the
apparent size of the galaxy and its inclination. In addition, the
spectral resolution of SDSS spectra is $\sim 100\ {\rm km s^{-1}}$,
making it difficult to reliably measure $\sigma_*$ for SMBHs with
$M_{BH} < 10^7 M_{\odot}$. Another concern is
that the local BHMF is derived by assuming that the 
$M_{BH}$--$\sigma_*$ or $M_{BH}$--$L_{sph}$ relations are single
power-laws with a constant scatter in $M_{BH}$ at fixed $\sigma_*$ or
$L_{sph}$. However, recent work has shown these assumptions to be
incorrect. For one, the $M_{BH}$--$\sigma_*$ and $M_{BH}$--$L_{sph}$
relations diverge at the high $M_{BH}$ end, which Lauer et al. \cite{lauer07}
suggest implies that the $M_{BH}$--$\sigma_*$ relation is not a single
power-law. This divergence creates an inconsistency in the BHMFs
derived from these two scaling relationships
\cite{lauer07,tundo07}. Similarly, the $\sigma$--$L$ relationships for the SDSS and
dynamical $M_{BH}$ SMBH samples are inconsistent, suggesting a
possible selection bias in the estimated BHMFs \cite{yu02,bernardi07}.  The scatter in the $M_{BH}$--$\sigma_*$
relation is larger for spirals \cite{gult09b}, and appears to
increase at low $M_{BH}$ such that most SMBHs lie below the
$M_{BH}$--$\sigma_*$ relation, \cite[e.g.]{greene10}. Several authors have found differences in
the slope and scatter of the scaling relations for pseudobulges
\cite{hu08,greene08,graham09}; however, it is unclear that this
result is due to differences in the perceived bulge velocity
dispersions for bulges as compared to pseudobulges or due to 
different scaling relationships. Kormendy, Bender, \& Cornell \cite{korm11} argue that
$M_{BH}$ does not correlate with galaxy disks, and only correlates
weakly, if at all, with pseudobulges. Although there is still much
that we do not understand about the $M_{BH}$ and host galaxy scaling
relationship, these recent results suggest that the scaling
relationships are not a single power-law with constant intrinsic
dispersion in $M_{BH}$, representing a significant source of
systematic uncertainty in the estimated local BHMF, especially at the
low-mass end. 

  \section{Black Hole Mass Functions Derived from the Local Mass
    Function and the AGN Luminosity Function}

  \label{s-continuity}

  By employing the argument of Soltan \cite{soltan82}, numerous studies have
  attempted to estimate the BHMF at a variety of redshifts by
  comparing the accreted mass distribution implied by the quasar
  luminosity function with the local BHMF
  \cite{yu02,yu04,marc04,merloni04,hopkins07,merloni08,cao08,cao10}. These
  methods employ a continuity equation describing the evolution of the
  number density of SMBHs \cite{cav71,small92}: 
  \begin{equation}
    \frac{\partial \phi_M(M_{BH},t)}{\partial t} + \frac{\partial
      \phi_M(M_{BH},t) \langle \dot{M}(M_{BH},t) \rangle}{\partial
      M_{BH}} = n_{merge}(M_{BH},t). 
    \label{eq-cont}
  \end{equation}
  Here, $\langle \dot{M}(M_{BH},t)\rangle$ is the average growth rate
  of SMBHs as a function of $M_{BH}$ and cosmic age, $t(z)$, and
  $n_{merge}(M_{BH},t)$ is the rate at which the number density of
  SMBHs changes due to mergers of black holes, or ejections of black
  holes from their host galaxies due to gravitational
  recoil. Technically, 
  $n_{merge}(M_{BH},t)$ can also include a contribution from SMBHs
  which are created, but this has not been thought to occur over the
  redshift range in which Equation (\ref{eq-cont}) is typically
  applied, i.e., $z \lesssim 5$. Because the merger rate of black
  holes is currently unknown, many studies that have employed Equation
  (\ref{eq-cont}) set $n_{merge}(M_{BH},t) = 0$.  

  Under the assumption that SMBHs grow during phases of AGN activity,
  AGN demographics in combination with the local BHMF may be used to
  compute $\phi_M(M_{BH},t)$. This is because the AGN luminosity
  function maps the accretion history onto SMBHs, and the local BHMF
  acts as a boundary condition on Equation (\ref{eq-cont}); it is
  also possible in principle to include the BHMF for AGN, which provides more
  information. Studies that have used Equation (\ref{eq-cont}) to
  estimate the BHMF generally fall into two categories: those that
  assume an AGN lightcurve, and those that employ the BHMF of AGN. We
  discuss each of these seperately. 

  \subsection{Methods that Assume an AGN Lightcurve}

  \label{s-lcurve}

  Most authors employing Equation (\ref{eq-cont}) have assumed a
  parameteric form for $\langle \dot{M}(M_{BH},t) \rangle$. The
  accretion rate is related to the bolometric luminosity output of the
  accretion flow onto the SMBH as $L = \epsilon_r \dot{M}_{acc} c^2$,
  where $\epsilon_r$ is the radiative efficiency of the accretion
  flow, $\dot{M}_{acc}$ is the accretion rate of matter onto the SMBH,
  and $c$ is the speed of light. The growth rate of the SMBH is
  $\dot{M} = (1 - \epsilon_{r}) \dot{M}_{acc}$, due to the fact that a fraction
  $\epsilon_r$ of accreted mass is radiated away as energy. Making
  this substitution, the continuity equation becomes 
  \begin{equation}
    \frac{\partial \phi_M(M_{BH},t)}{\partial t} + \frac{1 -
      \epsilon_r}{\epsilon_r c^2}\frac{\partial \phi_M(M_{BH},t)
      \langle L(M_{BH},t) \rangle}{\partial M_{BH}} = 0, 
    \label{eq-cont2}
  \end{equation}
  where we have ignored mergers of SMBHs. Equation (\ref{eq-cont2})
  shows that it is possible to calculated the BHMF at an time given
  the local BHMF, an assumed average accretion flow lightcurve as a
  function of $M_{BH}$, $\langle L(M_{BH},t) \rangle$, and an assumed
  radiative efficiency. Because $\phi_M(M_{BH},z)$ and $\langle
  L(M_{BH},t(z)) \rangle$ imply a luminosity function, the local BHMF
  and AGN luminosity function can be used to place constraints on
  $\epsilon_r$ and $\langle L(M_{BH},t) \rangle$. This means that, in
  practice, one also has to assume a bolometeric correction, which
  itself likely
  depends on both black hole mass \cite{kelly08}
  and $L / L_{Edd}$ \cite{vasud07,vasud09}. In addition, an
  estimate of $\langle L(M_{BH},t) \rangle$ also enables one to
  estimate the lifetime and duty cycle of AGN activity, modulo some
  luminosity-dependent definition of an AGN; note that the AGN duty
  cycle defines the fraction of SMBHs that are `active' at a given
  $M_{BH}$ and $z$. 

A variety of lightcurve models have been used when employing Equation
(\ref{eq-cont2}) to reconstruct the evolution of the BHMF. The
simplest model is that where SMBHs spend a fraction of their time
radiating at a constant Eddington ratio, and spend the remainder of
their time in quiescence. The free parameters in this model are the
Eddington ratio, AGN lifetime or duty cycle, and radiative
efficiency. This model has been used by
\cite{sal99,yu02,marc04,shank04,shank09}\footnote{Technically, \cite{sal99} assumed that the
  Eddington ratio was a weakly increasing function of luminosity.} to study the
build-up of the local black hole mass function, although \cite{shank09} also considered models where the average accretion rate relative to Eddington falls off toward lower $z$ and higher $M_{BH}$. Raimundo \& Fabian \cite{raim09} employed a variation on the constant $L / L_{Edd}$ models, assuming three different populations of AGN with their own Eddington ratio: a population of obscured low $L / L_{Edd}$ AGN, a population of obscured AGN with higher $L / L_{Edd}$, and a population of unobscured AGN. Yu \& Lu \cite{yu08} modeled
the quasar lightcurve as radiating at the Eddington limit for a period
of time, and then transitioning into a power-law decay. Cao \cite{cao10}
also modeled the quasar lightcurve as undergoing a power-law
decay. Lightcurves undergoing a power-law decay arise from
self-regulation models, and describe the evolution of the lightcurve
after black hole feedback unbinds the accreting gas, therefore
quenching its fuel supply. The power-law decay occurs either as a
result of evolution of a blast-wave \cite{stochacc,faintend} or from viscous
evolution of the accretion disk \cite{yu05,king07}.

\begin{figure}[t]
  \includegraphics[scale=0.5,angle=90]{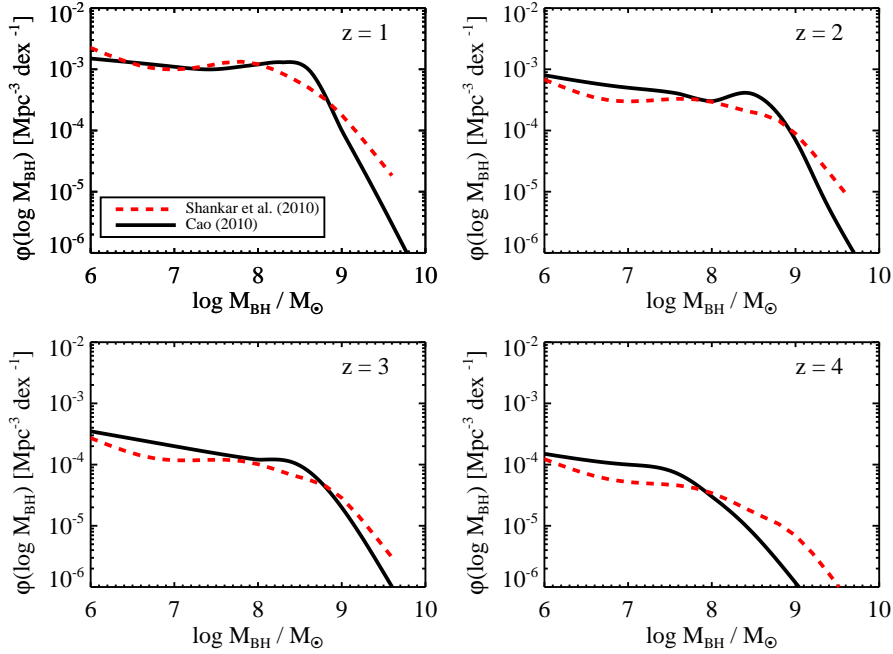}
  \caption{Comparison of two recently estimated BHMFs, calculated by
    Shankar et al. \cite{shank09} (red dashed line) with that
    calculated by Cao \cite{cao10} (solid black line). Both BHMF were
    estimated by assuming a quasar lightcurve, where Shankar et
    al. \cite{shank09} used a step function model while Cao
    \cite{cao10} assumed a power-law decay. Despite the two different
    models, the BHMFs are similar at a variety of redshifts, except at
    possibly the high mass end.} 
  \label{f-lcurve}
\end{figure}

In Figure \ref{f-lcurve} we compare the BHMF calculated by Shankar et
al.\cite{shank09} with that calculated by Cao \cite{cao10}. For
Shankar et al. \cite{shank09} we show their reference model, which
assumes a radiative efficiency of $\epsilon_r = 0.065$, an accretion
rate relative to Eddington of $\dot{M} / \dot{M}_{Edd} = 0.6$, and
that half of all SMBHs are active at $z = 6$. We show the model from
Cao \cite{cao10} which assumes a radiative efficiency of $\epsilon_r =
0.11$ and a quasar lifetime of $7.5 \times 10^8\ {\rm yr}$, as it
better matches the Shankar et al. \cite{shank09} estimates. The two
estimates of the BHMF agree fairly well, despite the different quasar
lightcurve models.

In general, most of the studies that have used Equation
(\ref{eq-cont2}) in combination with an assumed quasar lightcurve have
concluded the following:
\begin{itemize}
\item
  Most SMBH growth occurs in periods when the quasar is
  radiating near the Eddington limit.
\item 
  Most, if not all, of the
  local black hole mass function can be explained as the relic of
  previous AGN activity, implying that mergers of SMBHs are not
  important for building up the local mass function.
\item
  SMBH growth is anti-hierarchical, with the most massive black holes
  growing first. This has also been termed `down-sizing' of active
  SMBHs.
\item
  The lifetime of AGN activity is $\sim {\rm\ a\ few} \times 10^8 yr$.
\item
  Most SMBHs have non-zero spin, as implied by inferred radiative
  efficiencies of $\epsilon_r \gtrsim 0.06$.
\end{itemize}
However, while Equation (\ref{eq-cont2}) has proven to be an
important tool for studying SMBH growth, and estimating the black hole
mass function, it must be kept in mind that use of Equation
(\ref{eq-cont2}) often entails some strong assumptions. These methods
rely on the assumed form of the quasar 
lightcurve, distribution of radiative efficiencies, and bolometeric
corrections, all of which are subject to considerable
uncertainty. Moreover, in general these methods also rely on an estimate of the
local black hole mass function, which, as discussed in \S~\ref{s-local}, is
itself subject to considerable uncertainty. Indeed, there is a strong degeneracy between the estimated radiative efficiency of accretion and the normalization of the local BHMF, and therefore the uncertainty in $\epsilon_r$ is linearly proportional to that in the normalization (or integral) of the local BHMF. All of these issues have
the potential to introduce systematic error into methods based on
Equation (\ref{eq-cont2}), and further work is needed in reducing
these systematics.

  \subsection{Methods that Include the Distribution of Active
    Supermassive Black Holes}

  \label{s-fp}

  An alternative to the methods described in \S~\ref{s-lcurve} is to
  estimate the average value of the accretion rate onto SMBHs directly
  from the observational data. This avoids the issue of
  assuming a form for the quasar lightcurve, as instead $\langle
  L(M_{BH},t)\rangle$ is derived directly from the estimated
  distribution of $L / L_{Edd}$. Techniques based on this approach
  require a means of linking the mass function of active SMBHs to
  observational quantities, which is done via scaling
  relationships. This was the approach of Merloni \cite{merloni04} and
  Merloni \& Heinz \cite{merloni08}, who employed the black hole `fundamental plane'
  (BHFP) \cite{merloni03,falcke04}. 

\begin{figure}[t]
  \includegraphics[scale=0.4,angle=0]{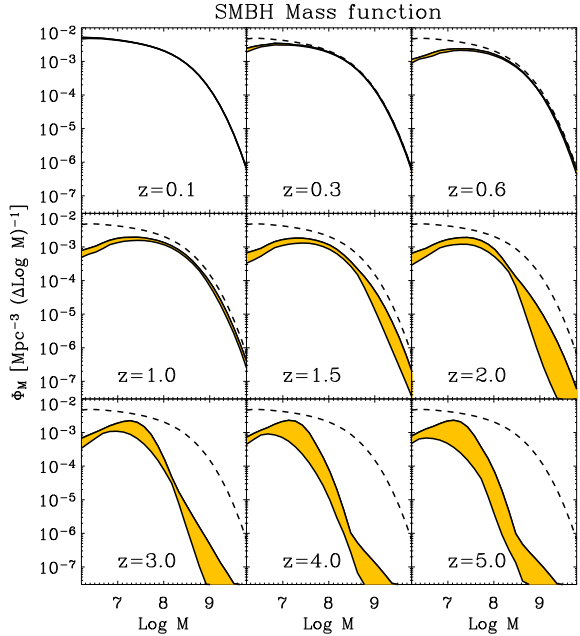}
  \caption{Figure 5 from Merloni \& Heinz \cite{merloni08}, showing
    the redshift evolution of their estimated BHMF. The dashed line is
    the local BHMF and the shaded regions reflect the uncertainty in
    the BHMF that is due to uncertainties in the AGN luminosity
    function. The high mass end of their estimated BHMF is built up
    faster than the low mass end, a phenomenon that has been called
    `downsizing'.}
  \label{f-mh08_bhmf}
\end{figure}

  The BHFP is a scaling
  relationship between $M_{BH}$, radio luminosity, and X-ray
  luminosity, that exists for low-accretion rate black holes (i.e., $\dot{M} / \dot{M}_{Edd} \lesssim 0.01$), extending
  from galactic black holes to supermassive ones. The BHFP likely
  reflects the connection between $M_{BH}$ and the conversion of the
  accretion flow into radiative energy and jet power. It, in
  principle, enables one to connect the radio and X-ray luminosity
  functions to the mass function of active SMBHs. Having obtained a
  distribution of $M_{BH}$ and X-ray luminosity at a given redshift for the active SMBH population, Merloni \cite{merloni04}
  and Merloni \& Heinz \cite{merloni08} then convert this to a joint distribution of
  $M_{BH}$ and $\dot{M}_{acc}$ assuming a conversion from X-ray luminosity
  to $\dot{M}_{acc}$ which depends on the Eddington ratio. The joint
  distribution of $M_{BH}$ and $\dot{M}_{acc}$ at a given redshift for active SMBHs therefore
  enables calculation of the average growth rate $\langle \dot{M}(M_{BH},t(z))
  \rangle$, which can then be combined with the continuity equation to
  calculate the black hole mass function at the next redshift. Their estimated BHMF is shown in Figure \ref{f-mh08_bhmf}, which is a recreation of their Figure 5. Similar to
  methods based on assuming a quasar lightcurve, Merloni \& Heinz \cite{merloni08}
  concluded that SMBHs grow anti-hierarchically; however, in contrast
  to the lightcurve methods, Merloni \& Heinz \cite{merloni08} concluded that most
  SMBHs have low spin as inferred from their derived radiative
  efficiency. In addition, Merloni \& Heinz \cite{merloni08} concluded that the distribution of
  SMBH accretion rates is broad, and that most SMBH growth occurs during
  a radiatively efficient accretion mode.

  The method of estimating the BHMF from the BHFP developed by
  Merloni \cite{merloni04} and Merloni \& Heinz \cite{merloni08} has the advantage that it
  derives the distribution of accretion rates empirically. However, there are
  also disadvantages to this approach. The uncertainties regarding the
  bolometeric correction, estimation of the local BHMF, and radiative
  efficiency also apply to the BHFP method as well. Moreover, as
  discussed in Merloni \& Heinz \cite{merloni08}, the BHFP
  is only defined for low-accretion rate objects, i.e., objects with
  $L / L_{Edd} \lesssim 10^{-2}$. Merloni \& Heinz \cite{merloni08} extrapolate the
  BHFP to higher accretion rates, after rescaling the normalization to
  ensure that the radio luminosity is weak for AGN in the radiatively
  efficient mode (those objects with $L / L_{Edd} \gtrsim 10^{-2}$ and
  lacking a jet). Unfortunately, the AGN in the radiatively efficient
  mode make a significant contribution to the X-ray luminosity
  function, from which $\langle \dot{M}(M_{BH},t)$ is
  derived. Moreover, most studies, including those based on the BHFP
  have concluded that most SMBH 
  growth occurs at $L / L_{Edd} \gtrsim 10^{-2}$, which corresponds to
  the radiatively efficient mode. Because the radiatively efficient
  mode also corresponds to the regime of largest systematic
  uncertainty for the BHFP, there is the potential
  for significant systematic error in estimating the BHMF based on
  the BHFP, as well as in estimating the primary mode of SMBH
  growth. There is thus a need for further improvement to our
  understanding of the scaling relationships involving $M_{BH}$ and
  the AGN SED.

  \section{Black Hole Mass Functions of AGN}

  \label{s-agn}

  Thus far, we have focused on methods for estimating the mass
  function of all SMBHs. In this section we will describe methods for
  estimating the BHMF for those SMBHs in AGN, and the results that
  have come from the application of these methods.

  \subsection{Methods Based on Scaling Relationships Involving the
    Broad Emission Lines}

  \label{s-broad}

  The steady improvement in reverberation mapping of AGN
  \cite{peter04,bentz_rm} has revealed a correlation between
  the luminosity of AGN and the broad line region radius
  \cite{kaspi05,bentz_rl}. It is therefore possible, in principle, to
  obtain an estimate of $M_{BH}$ for broad line AGN (BLAGN) by
  combining a luminosity-based estimate of the broad line region size
  with an estimate of the velocity dispersion of the broad line region
  gas obtained from the width of the broad emission lines
  \cite{wandel99}. These virial mass estimates are then
  calibrated to the estimates of $M_{BH}$ obtained from reverberation
  mapping, which themselves are calibrated to be consistent with the
  local $M_{BH}$--$\sigma_*$ relationship
  \cite{onken04,woo10}. Currently, calibrations exist for H$\alpha$
  \cite{greene05}, H$\beta$ \cite[e.g.]{vest06}, Mg II
  \cite{mclure_jarvis02,vest09,shen11}, and C IV \cite{vest06}. The
  statistical scatter in the virial mass estimates is currently
  estimated to be $\sim 0.4$ dex \cite{vest06}, although there are
  indications that the scatter may be smaller, at least for the most
  luminous quasars
  \cite{koll06,shen08,stein10,kelly10,shen_kelly11}. Moreover, it should
  be noted that the calibration for Mg II is obtained by enforcing
  consistency in the mean values of the Mg II mass estimator and the H$\beta$ and C
  IV ones, and therefore there is currently no direct estimate of the
  statistical scatter in Mg II-based virial mass estimates.  In
  contrast, the amplitudes of the statistical scatter for H$\beta$ and
  C IV are estimated by comparing mass estimates derived from these
  lines with the masses derived from reverberation mapping \cite{vest06}. Although
  there is currently very little reverberation mapping data for C IV,
  the estimate of the dispersion in the C IV-based mass estimates
  should not be biased so long as the masses based on reverberation mapping are
  reliable estimates of the true $M_{BH}$, regardless of which
  emission line was used in the reverberation mapping campaign.

Early estimates of the mass function of SMBHs in BLAGN were obtained
by binning up the virial mass estimates and applying a $1 / V_{max}$
correction \cite{wang06,greene07,vest08,vest09}, a technique borrowed
from luminosity function estimation. Greene \& Ho \cite{greene07} estimated the
local BHMF for BLAGN from the SDSS DR4, while Vestergaard et al. \cite{vest08} estimated
the BHMF for BLAGN over $0.3 < z < 5$ using the uniformly-selected
quasar sample from the SDSS DR3 \cite{rich06}. Vestergaard \& Osmer \cite{vest09}
estimated the BHMF for the brightest BLAGN using objects from a
variety of surveys, as their sample was designed to complement the
uniformly-selected SDSS DR3 sample. Unfortunately, as discussed in
\S~\ref{s-complications}, this method of binning up the mass estimates
suffers from biases due to the large statistical scatter in the virial
mass estimates, and due to the inability of a luminosity-based $1 /
V_{max}$ correction to correct for incompleteness in $M_{BH}$. Subsequent attempts
have further improved in their methodology, providing more accurate
BHMFs.  

\begin{figure}[t]
  \includegraphics[scale=0.45,angle=90]{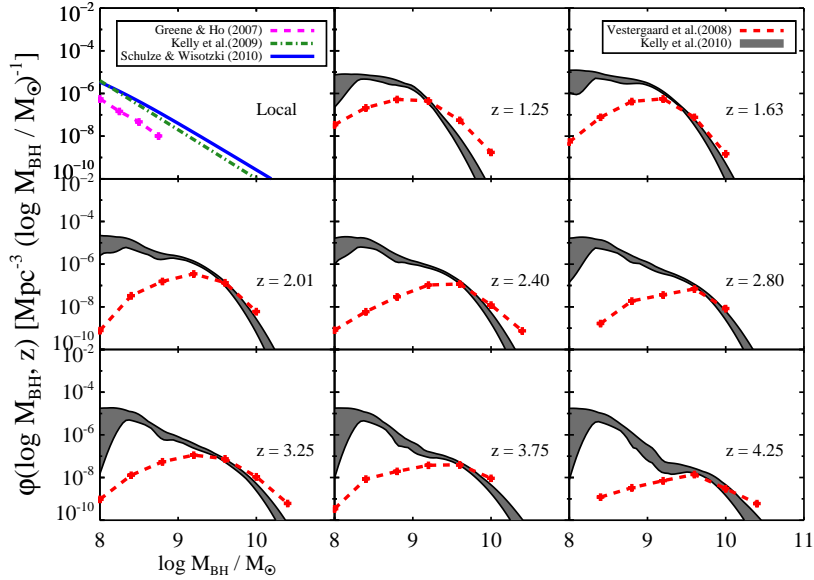}
  \caption{Broad-line AGN BHMFs at a variety of redshifts. Shown are
    the local BHMF estimated by Greene \& Ho \cite{greene07} (dashed
    magenta line), Kelly et al. \cite{kvf09} (dot-dashed green line),
    and Schulze \& Wisotzki \cite{schulze10} (solid blue line). Also
    shown are the $z > 1$ BHMFs estimated by Vestergaard et
    al. \cite{vest08} (dashed red line) and Kelly et
    al. \cite{kelly10} (shaded region), where the shaded region for
    the Kelly et al. \cite{kelly10} estimate defines an approximate
    $95\%$ confidence region. The BHMFs estimated by Greene \& Ho
    \cite{greene07} and Vestergaard et al. \cite{vest08} are
    flux-limited BHMFs, as they did not fully correct for
    incompleteness in $M_{BH}$.} 
  \label{f-blqso_bhmf}
\end{figure}

Shen et al. \cite{shen08} employed a forward-modeling approach where the mass
function and Eddington ratio distribution were estimated by matching
the observed distribution of mass estimates and luminosity to that
implied by the model BHMF and Eddington ratio distribution. Their
method accounts for incompleteness and the statistical scatter in the
mass estimates, but lacked statistical rigor in that the matching was
done visually. Schulze \& Wisotzki \cite{schulze10} employed a maximum-likelihood
technique for estimating the local BHMF for BLAGN. Their method
corrects for incompleteness in $M_{BH}$ but does not correct the BHMF
for the broadening caused by the statistical scatter in the virials
mass estimates. Kelly et al. \cite{kvf09} developed a Bayesian method that
corrects for both the statistical scatter in the mass estimates and
incompleteness, and used their method to estimate the local BHMF of
BLAGN from the Bright Quasar Survey \cite{bqs}. Kelly et al. \cite{kelly10} used
the method of \cite{kvf09} to estimate the BHMF of BLAGN at $1 < z <
4.5$ from the mass estimates in the SDSS DR3 quasar sample
\cite{vest08}. The BLAGN BHMFs from a variety of studies are compiled in Figure \ref{f-blqso_bhmf}, showing the evolution of the BHMF from the local universe out to $z = 4.5$. More recently, Shen \& Kelly \cite{shen_kelly11} extended the Bayesian method of \cite{kvf09} to include a possible
luminosity-dependent bias in virial mass estimates derived from the
emission line $FWHM$, the existence of which was suggested by
Shen \& Kelly \cite{shen_kelly10}. Shen \& Kelly \cite{shen_kelly11} applied their method to the
SDSS DR7 uniformly-selected quasar sample, independently estimating
the BHMF and Eddington ratio distribution in different redshifts
bins.

Similar to the methods based on the continuity equation,
investigations of the BHMF for BLAGN have found evidence for the
anti-hierarchical growth of SMBHs, i.e., cosmic `down-sizing' of BLAGN
activity. The inferred Eddington ratio distributions are
wide, and the density of SMBHs continues to increase toward Eddington
ratios which are below the survey completeness limit. In
addition, Kelly et al. \cite{kelly10} used the BLAGN BHMF to estimate the lifetime
of broad line quasar activity to be $t_{BL} \sim 150$ Myr among SMBHs
with $M_{BH} \sim 10^9 M_{\odot}$, which is similar to quasar
lifetimes inferred from the continuity equation. Kelly et al. \cite{kelly10} also
used their estimated BHMF to estimate the maximum mass of a SMBH to be
$M_{BH} \sim 3 \times 10^{10} M_{\odot}$, which is in agreement with
theoretical expectations \cite{priya09,sijacki09}. 

Mass functions estimated from scaling relationships for BLAGN have the
advantage that they are derived from estimates of $M_{BH}$ that are
obtained for individual sources, providing a more `direct' estimate of
the mass function than those based on the continuity
equation. However, they have the disadvantage that they are only
available for a subset of the AGN population, which itself is only a
subset of the SMBH population. This complicates comparison with other
SMBH mass functions, as the fraction of AGN with broad emission lines
is poorly constrained, especially as a function of mass. This being
said, BHMFs of BLAGN represent 
a subset of SMBHs that are actively growing at the time that they
are observed, and, as the aforementioned studies have demonstrated,
their mass function still contains important information on SMBH growth.  

As with all methods of BHMF estimation, the virial mass estimates and
the mass functions derived from them still suffer from
systematics. First, there is the usual problem of calculating a
bolometeric correction, although this only affects the estimated
Eddington ratio distribution, and not the BHMF. Second, there are a
few concerns with the virial mass estimates which could introduce
systematic error; some of these have been discussed by Greene \& Ho
\cite{greene09}. For one, most of the reverberation mapping data is
only available for the H$\beta$ line. Because of the limited Mg II
data, the Mg II scaling relationship is in general not calibrated using objects
with black hole mass estimates from reverberation mapping. There may
be systematic effects with luminosity or Eddington ratio when using
the $FWHM$-based scaling relationships \cite{collin06,shen_kelly11},
possibly due to a dependence of the broad line region structure on
these quantities. Systematic effects on broad line region geometry, which can effect the inferred velocity dispersion, are
a particular concern for C IV, which is thought to arise in an
accretion disk wind \cite{richards11}. Along these lines, unaccounted
for radiation pressure on broad line clouds may also bias the virial
masses, especially among those AGN radiating near the Eddington limit,
\cite{marconi08}; however, its importance is still debated
\cite{netzer09,marconi09,netzer10}. In addition, the reliability of
line width measurements can rapidly deteriorate for low $S / N$ data
\cite{denney09}. And finally, the BLAGN virial mass estimates are
calibrated to the reverberation mapping derived masses, which
themselves are calibrated to lie on the local $M_{BH}$--$\sigma_*$
relationship. Most of the AGN that are used to calibrate the
reverberation mapping masses to the $M_{BH}$--$\sigma_*$ relationship
have lower masses and are hosted by late type galaxies, for which
there is evidence that the $M_{BH}$--$\sigma_*$ relationship begins to
break down \cite{greene10}. Greene et al. \cite{greene10} argue that the normalization of the scaling
relationships inferred when limiting the calibration to low mass SMBHs
hosted in late type galaxies may be about a factor of $\sim 1.5$ lower
than that used for the current broad line mass estimates
\cite{greene10}. However, dynamical mass estimates exist for two
reverberation mapped AGN: NGC 3227 \cite{davies06,hicks08} and NGC
4151 \cite{onken07,hicks08}. In both cases the masses derived from
dynamical modeling and reverberation mapping agree, so it is unclear
if a smaller scaling factor is needed for late-type galaxies. These
issues show that there are still many remaining questions
regarding virial masses, highlighting the need for further study using
high-quality reverberation mapping data.

  \subsection{Other Methods for Estimating the Black Hole Mass
    Function of AGN}

  \label{s-other}

  Before broad line mass estimates, there were two earlier attempts at
  estimating the BHMF for AGN, which we briefly mention
  here. Siemiginowska \& Elvis \cite{siem97} and 
  Hatziminaoglou, Siemiginowska, \& Elvis \cite{hatz01} used a model for the AGN lightcurve arising due to thermal-viscous
  accretion disk instabilities \cite{siem96} to calculate the
  expected distribution of luminosity at a given black hole
  mass. Based on this calculated distribution, they used the quasar
  luminosity function to constrain the quasar black hole mass
  function. Siemiginowska \& Elvis \cite{siem97} found evidence for SMBH downsizing in AGN,
  consistent with later work.

  Franceschini et al. \cite{franc98} found a tight correlation between $M_{BH}$ and the
  total radio power observed in a sample of local galaxies. They then
  used their empirical relationship to estimate the local BHMF derived
  from the local radio luminosity function of galaxies. While many of
  the objects in their sample are not considered AGN in the
  traditional sense, Fraceschini et al. \cite{franc98} argue that this correlation is a
  signature of an advection-dominated accretion flow, thought to
  dominate at low accretion rates relative to Eddington. Therefore,
  while these SMBH may not be `active' in the quasar sense, the
  determination of their mass function relies on radio emission from
  the SMBH accretion flow, so this method may still be considered a
  method for estimating the BHMF for active SMBHs. Franceschini et
  al. \cite{franc98}
  compared their BHMF to models of AGN activity and found that it was
  inconsistent with AGN activity being continuous and long-lived, but
  consistent with AGN activity being transient and possibly
  recurrent. 

  \section{Theoretical Models for Black Hole Mass Functions Across
    Cosmic Time}

  \label{s-theory}

  There have been numerous theoretical models for the formation and
  growth of supermassive black holes, and coevolution with their
  host galaxies. Understanding this formation, growth, and coevolution
  is one of the current most important outstanding issues
  in extragalactic astrophysics. Because the black hole mass function
  provides a census of the SMBH population and its evolution, it is
  one of the most fundamental observational quantities available for
  constraining models of SMBH formation and growth. As such, many
  theoretical investigations have predicted a BMHF for comparison with
  the empirical BHMF. In this section we review some of the models for
  SMBH formation and growth. There have been numerous theoretical
  models for SMBH growth and formation, and it is beyond the scope of
  this primarily empirically-focused review to review all of them;
  instead, we focus on those theoretical models that predict a
  BHMF.

  \subsection{Modeling the Coevolution of SMBHs and Galaxies:
    Predicted BHMFs}
  
  \label{s-models}

  Early models for the coevolution of SMHBs and galaxies linked the
  growth of black holes to the properties of host dark
  matter halos, with periods of SMBH growth occuring in quasar phases
  initiated by mergers. In general, early studies that predicted a
  BHMF used various
  perscriptions to relate $M_{BH}$ to the mass of the host halo
  \cite{haeh93,kauff00, hoso02,cav02,enoki03}. More recent models for the coevolution of SMBHs and galaxies have
  incorporated AGN feedback from the SMBH. In addition, the
  availability of empirical BHMFs have enabled modelers to compare
  their more recent models with observational data. In general, the models are
  qualitatively in agreement with the empirical results, in that they
  are able to match the local BHMF fairly well and predict downsizing
  of SMBHs. However, considering the current systematic and
  statistical uncertainties in the empirical results, it is difficult
  to place rigorous empirical constraints on the models such that
  certain models may be ruled out. Because of this, we simply
  summarize some of the different recent models that have been developed
  which predict the BHMF.

 Granato et al. \cite{granato04} developed
  a model incorporating feedback from AGN and supernovae, where the
  feeding of the SMBH is driven by stellar radiation drag on
  gas. Their predicted local BHMF agrees with that estimated by
  Shankar et al. \cite{shank04}. Cattaneo et al. \cite{catt05} used halo merger trees constructed
  from $N$-body simulations to track the growth of SMBHs. In their
  model the black hole fueling rate was proportional to the star
  formation rate of the host galaxy burst component and the density of
  the cold gas in the starburst component. Their model predicted SMBH
  downsizing, with the most massive part of the BHMF being built up
  first, in agreement with the subsequent empirical
  studies. 

  Hopkins et al. \cite{hopkins08} describe a model for the coevolution of
  SMBHs and galaxies whereby all major mergers of gas-rich galaxies
  trigger a quasar. In this model the final black hole mass is assumed
  to be on average proportional to the host spheroidal mass, in
  agreement with the local scaling relationships between SMBHs and
  their host galaxies. Hopkins et al. \cite{hopkins08} estimated  the merger rate of
  gas-rich galaxies by combining theoretical constraints of the halo
  and subhalo mass functions with empirical constraints on halo
  occupation models. Their model also predicts SMBH downsizing, and
  their predicted BHMF matches the local BHMF derived by
  Marconi et al. \cite{marc04}. Similarly, Shen \cite{shen09} also assumed that quasars
  are triggered by major mergers of gas-rich galaxies, with the SMBHs
  growing via accretion in these quasar phases. Shen \cite{shen09} used a
  halo merger rate based on theoretical expectations from $N$-body
  simulations, and assumed a universal quasar lightcurve shape having
  an exponential increase followed by a power-law decay (see
  also \cite{yu08}). The BHMF predicted by Shen \cite{shen09} broadly agrees
  with the local one estimated by Shankar et al. \cite{shank09} and predicts that
  most SMBHs with $M_{BH} > 3 \times 10^8 M_{\odot}$ were in place by
  $z = 1$, but only $50\%$ of them were assembled by $z = 2$. 

  Most recently, Fanidakis et al. \cite{fan11a,fan11b} extended the model of
  \cite{bower06}, which includes AGN feedback, to also follow the
  spin distribution of SMBHs. In their model SMBHs are fueled through
  accretion of cold gas from mergers, disk instabilities, and cooling
  flows from hot halos. However, the inclusion of SMBH spin enabled
  them to include different radiative efficiencies, which dictates how
  much accreted material actually grows the black hole, and to provide
  an improved model for the amount of mechanical feedback imparted through an AGN
  jet, both of which depend on the spin of the black hole. Their model
  predicts that the present-day Universe is dominated by SMBHs with
  $M_{BH} \sim 10^7$--$10^8 M_{\odot}$, and that the BHMF at $M_{BH} >
  10^9 M_{\odot}$ was largely built up at $z < 2$ due to an increase
  in both lower accretion rate `radio-mode' growth and mergers of
  SMBHs.

  Almost all models for the cosmological coevolution of SMBHs and
  galaxies that predict a BHMF have been of an analytical or semi-analytical nature. An
  exception is the study done by Di Matteo et al. \cite{dimatteo08}, who present the
  results from cosmological hydrodynamic simulations of the
  $\Lambda$CDM model that follow the growth of galaxies and SMBHs,
  including their feedback processes, at $z > 1$. Direct cosmological simulations
  such as these should, in principle, provide the most accurate results as to the
  predicted BHMF, and for identifying the relevant physical processes
  that are important in shaping the BHMF. However, current
  cosmological hydrodynamic simulations suffer from the fact that they cannot resolve processes on physical scales
  corresponding to the SMBH accretion flow. In fact,
  Di Matteo \cite{dimatteo08} use a gravitational softening length of $\epsilon = 2.73h^{-1}$ kpc. Instead,
  Di Matteo \cite{dimatteo08} employ a subresolution model where the accretion
  onto the SMBH is estimated using a Bondi-Hoyle-Lyttleton
  parameterization \cite{hoyle39,bondi44,bondi52} with a correction factor to account
  for the fact that the Bondi radius is not resolved. They assume a radiative feedback energy
  efficiency of $5\%$ \cite{dimatteo05}, which is the only free
  parameter in their model and required in order to match the normalization of the observed local $M_{BH}$--$\sigma_*$ relationship. Their calculated BHMF at $z = 1$ matches
  the local BHMF for $M_{BH} > 2 \times 10^8 M_{\odot}$. In addition,
  Di Matteo \cite{dimatteo08} also find downsizing in their model, in agreement
  with observations, with the high mass end of the BHMF being largely
  in place by $z \sim 2$.

\begin{figure}[t]
  \includegraphics[scale=0.45,angle=90]{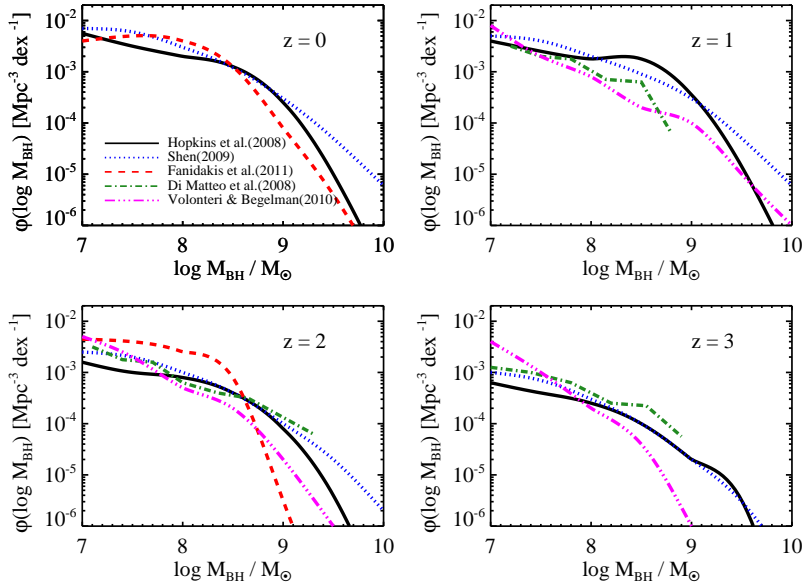}
  \caption{Compilation of BHMFs predicted by several recent models for SMBH formation and growth. Shown are the BHMFs predicted by Hopkins et al. \cite{hopkins08} (solid black line), Shen \cite{shen09} (dotted blue line), Fanidakis et al. \cite{fan11a} (dashed red line), Di Matteo et al. \cite{dimatteo08} (dot-dashed green line), and Volonteri \& Begelman \cite{vol10} (dashed-dotted-dotted-dotted magenta line). In general the number densities predicted by the models agree to within a factor of a few, although they diverge at $M_{BH} \gtrsim 10^9 M_{\odot}$. Some of these authors did not report BHMFs at each redshift shown, so we only show those available at each redshift.}
  \label{f-theory}
\end{figure}

In Figure \ref{f-theory} we compile predicted BHMFs from several recent models for SMBH formation and growth \cite{hopkins08,shen09,fan11a,dimatteo08,vol10}. In general, the models tend to agree to within a factor of a few with regards to the BHMF. However, they diverge at $M_{BH} \gtrsim 10^9 M_{\odot}$, where their predicted SMBH number densities can differ by over an order of magnitude. 

  \subsection{Modeling the BHMF of SMBH Seeds}

  \label{s-seeds}

  Recent work has made improvement to models for the BHMF by
  focusing on theoretical modeling of the distribution of seed
  SMBHs. The discovery of quasars at $z \approx 6$--$7$ with $M_{BH} \sim 10^9
  M_{\odot}$ \cite{jiang07,mortlock11} places strong constraints on the
  formation of SMBH seeds due to the very limited amount of time
  available at that redshift to grow SMBHs. Lodato \& Natarajan \cite{lodato07} derive
  the BHMF of SMBH seeds at $z \sim 15$ 
  that are the result of the collapse of pregalactic disks which have
  not yet been enriched by metals \cite{lodato06}. Black holes formed
  through such a mechanism have masses $M_{BH} \sim 10^5 M_{\odot}$,
  while black holes which are the remnants of Pop III stars have
  $M_{BH} \sim 10^3 M_{\odot}$. A similar seed black hole formation
  mechanism is through `quasi-stars' \cite{beg06,beg08,beg10}, which
  are also able to produce seed black holes with $M_{BH} \sim 10^5
  M_{\odot}$. 

  Volonteri, Lodato, \& Natarajan \cite{vol08} describe a model for the growth of SMBHs seeded
  according to the direct collapse model of Lodato \& Natarajan \cite{lodato06} with
  varying formation efficiencies. In addition, they also compared the
  results from this model using SMBHs seeded from
  Pop III remnants. Volonteri et al. \cite{vol08} grow SMBHs through major mergers,
  and force the black hole mass after the galaxy merger to scale with
  the circular 
  velocity of the host halo; additional growth is also provided
  through black hole mergers. Their merger trees are based on a Monte
  Carlo algorithm based on the extended Press-Schechter
  formalism. They find that most significant differences in the local BHMF with respect
  to black hole formation efficiency occur at $M_{BH} < 10^7
  M_{\odot}$, with the number density of SMBHs with $M_{BH} < 10^7
  M_{\odot}$ increasing with increasing formation
  efficiency. Volonteri \& Begelman \cite{vol10} performed a similar analysis as that of
  Volonteri et al. \cite{vol08} but instead used SMBH seeds formed via
  quasi-stars. The BHMFs calculated by Volonteri \& Begelman \cite{vol10} match those of
  Merloni \& Heinz \cite{merloni08} at the high mass end, at least at $z > 2$.

  Natarajan \& Volonteri \cite{nat11} used a growth and seeding model which is very similar to that
  employed by Volonteri et al. \cite{vol08}. However, they also predict the BHMF for
  broad line quasars, assuming that $20\%$ of quasars are
  unobscured. They compare the BHMF derived from their model at $1 < z
  < 4.5$ to the BHMF for broad line quasars reported by
  Kelly et al. \cite{kelly10}, and to the BHMF for all SMBHs reported by
  Merloni \& Heinz \cite{merloni08}. Natarajan \& Volonteri \cite{nat11} concluded that seeds from Pop III
  stars have difficulty reproducing the BLAGN BHMF, especially at high
  redshift, while seeds resulting from the direct collapse of
  pregalactic disks do better at fitting the high mass end of the
  BLAGN BHMF at $z > 2$.

  \section{Directions for Future Work and Improvement}

  \label{s-future}

  Before concluding this review, we present a discussion of possible future
  empirical and theoretical work relevant to BHMF studies. These
  include: 
  \begin{itemize}
  \item
    {\bf Better characterization of the SMBH-host galaxy scaling
      relationships}. Currently, the local BHMF is estimated from the
    distribution of host galaxy properties assuming that $M_{BH}$ has
    a constant log-normal scatter about a single-power law scaling
    relationship. As discussed in \S~\ref{s-local}, recent
    observations have provided reason to doubt this assumption,
    suggesting that the correlations break down at the highest and
    lowest masses. This will create biases in the BHMF 
    determined from the scaling relationships, which in turn will
    also affect the BHMF estimated from the continuity
    equation. Further direct $M_{BH}$ estimates from dynamical and
    kinematic modeling should be obtained for a variety of galaxy
    types, especially at the high and low $M_{BH}$ end. The next class
    of $25+$ m telescopes should provide a significantly improved
    picture of the scaling relationships, thus providing us with more
    accurate estimates of the local BHMF. 
  \item
    {\bf Improvements to techniques based on the continuity equation.}
    Most studies that have invoked the continuity equation to link the
    local BHMF to the AGN luminosity function have assumed a single
    radiative efficiency, which is equivalent to assuming a single
    black hole spin, and a universal AGN lightcurve. Neither of these
    assumptions are likely to be true, and improvements to this type
    of modeling should include a distribution of SMBH spin and AGN
    lightcurves. In addition, we need to better characterize the
    bolometeric corrections, which remain a significant source of
    systematic uncertainty. The continuity equation techniques should
    also be extended to map the evolution of the full joint
    3-dimensional distribution of black hole mass, accretion rate, and
    spin. While this will not necessarily have a direct effect on
    estimating the BHMF, it will provide insight into the dominant
    accretion modes experienced by active SMBH and into the dominant fueling mechanism
    for AGN activity, as the spin distribution traces the SMBH fueling
    history \cite{berti08}.
  \item
    {\bf Better characterization of scaling relationships involving $M_{BH}$ for AGN}. 
    The dominant scaling relationship for estimating $M_{BH}$ in AGN
    involves the broad emission lines. However, as discussed in
    \S~\ref{s-broad} a number of systematic uncertainties remain. In
    order to reduce these systematics, we need to better understand
    the broad line region geometry for the different emission lines,
    and how it scales with luminosity,
    which will provide us with a more accurate conversion from line
    width to velocity dispersion. Moreover, accurate
    characterization of the broad line region geometry should remove
    the need to calibrate the scaling relationships to the local
    $M_{BH}$--$\sigma_*$ relationship, which has its own set of
    systematics. Improvements to reverberation mapping campaigns and
    modeling
    \cite{horne04,bentz10,pancoast11,zu11,brewer11}, as well
    as increasing the number of AGN monitored for reverberation mapping, will be needed in order
    to really understand the systematics involved in the broad line
    mass estimates.

    There is also the need to better characterize the black hole
    fundamental plane. Because the BHFP describes how the emission
    mechanisms responsible for the radio and X-ray flux scale with
    $M_{BH}$, the BHFP coefficients depend on these emission
    mechanisms. However, these emission mechanisms depend on the
    geometry of the accretion state and the existence of a jet, which
    in turn depends on the accretion rate \cite{merloni05}, and
    therefore the BHFP coefficients will be different for different
    classes of AGN \cite{kord06,gult09a,plotkin11}. In
    particular, the BHFP is currently poorly constrained for
    `soft-state' galactic black holes and radio-quiet AGN. Therefore,
    to reduce the systematics
    involved with the BHFP it will be necessary to characterize the scaling relationships and
    their scatter for radio-quiet objects. A correlation between the radio and X-ray
    luminosity has been observed for radio-quiet objects
    \cite{laor08}, implying that a BHFP should also exist for
    these objects. In order to better characterize the BHFP for
    radio-quiet objects it will be necessary
    to obtain radio detections for a
    well-defined sample of radio-quiet AGN with reliable $M_{BH}$ estimates and
    X-ray detections.

    Finally, there has recently been the discovery of scaling
    relationships involving $M_{BH}$ and the optical
    \cite{kelly09,macleod10} and X-ray
    \cite{nik04,mchardy06,kelly11} variability properties of
    AGN. Mass estimators based on these scaling relationships
    have not been rigorously developed yet, nor have they seen
    widespread use. However, the existence of these scaling
    relationships implies that the variability properties may offer
    another avenue for estimating $M_{BH}$ and BHMFs, which may become
    increasingly valuable in the era of current and future large
    time-domain surveys, such as \emph{Pan-Starrs} and \emph{LSST}.
\item
  {\bf Understanding the redshift evolution of scaling relations.}
  From the theoretical point of view, it is clear that high-redshift
  scaling relations (or the lack thereof) between SMBH and their hosts
  provide unique and powerful constraints to models for AGN feeding and
  feedback, which cannot be otherwise distinguished (see e.g. Merloni et
  al.\cite{merloni10} and references therein).

  In practical terms, a better understanding of the evolution of scaling
  relations may also be very advantageous for BHMF studies. As we
  discussed above, current technique for BH mass estimation at z>0
  involve un-obscured, broad emission line QSOs. One can argue that, as
  long as we are restricted to just this class of QSOs, we will have to
  make critical assumptions about the properties of a significant part
  of the population to draw conclusions about the full BHMF (if we
  wanted, for example, to compare with "continuity-equation-based"
  methods). On the other hand, large multi-wavelength surveys do and
  will provide a wealth of information on the host galaxies of obscured
  AGN at high redshift, that represent the numerically dominant part of
  the growing black holes population
  \cite{brusa09,card10,xue10,aird11}. Therefore, if we
  had an independent way to put constraints on the nature of the BH-host
  relation for these objects, we could explore the uncharted territory
  of BHMF for obscured AGN (and for the entire population). Such an
  independent information could come, for example, from IR studies of
  broad emission lines which could act as probes of the BH potential
  less affected by obscuration. The first exploratory works pursuing
  this line of research have recently been published, e.g., \cite{alexander08,sarria10}.

  From the technical point of view, a lot of work of course is needed to
  better understand how reliable these estimators are. Another big
  "technical" challenge of all studies of the evolution of scaling
  relations, is the fact that they require a thorough assessment of the
  many observational biases one encounters in studying high redshift AGN
  and their hosts \cite{schulze11}.
\item
    {\bf Accounting for black hole kicks in theoretical models.}
    Most theoretical models for the BHMF do not include recoiling
    effects caused by the merger of two black holes. However, recent
    theoretical work on black hole recoils suggests that black holes
    can spend a significantly large enough amount of time offset from
    the central region of the host galaxy to alter their growth, thereby increasing
    the scatter about the scaling relationships and decreasing the final
    black hole mass
    \cite{vol07,blecha08,sijacki11,blecha11}. On the other
    hand, Volonteri, G{\"u}ltekin, \& Dotti \cite{vol10_eject} and
    Volonteri, Natarajan, G\"{u}ltekin \cite{vol11} find that ejected 
    SMBHs are rare at $z < 5$, especially for massive SMBHs,
    suggesting that accounting for ejected black holes will not make
    a significant difference in the BHMF. Further
    improvement in our understanding of
    the effects and frequency of black hole recoil will ensure the
    accurate implementation of black hole recoil into models for SMBH growth.
  \item
   {\bf Improvements to our understanding of AGN feedback.}   
   Most current theoretical models for SMBH growth involve AGN feedback, and assume a single
   efficiency for coupling feedback energy to the gas; this feedback
   efficiency is usally treated as a free parameter. An improved physical understanding of AGN
   feedback will improve theoretical models for the BHMF, as the feedback efficiency 
   affects the dynamics of the SMBH's fuel supply, and therefore the amount
   that the SMBH accretes as a function of redshift. Recent
   high-resolution hydrodynamic simulations in one dimension
   \cite{ciotti09,shin10,ciotti10} and two dimensions \cite{kuro09,ost10}
   have concluded that AGN feedback efficiency increases with the
   Eddington ratio, and that the values are below the value of $\sim
   5\%$ assumed in many current theoretical models for SMBH growth. Further
   improvements to simulations developed for studying AGN feedback will
   lead to a better physical understanding of AGN feedback, which will
   improve theoretical models for SMBH growth and the BHMF.

   On the observational side, future
   X-ray observations should provide considerable improvement in our
   understanding of AGN feedback. X-ray spectra are needed in order to
   determine the toal column density of the gas, and thus its kinetic
   energy flux, which can be compared to the energy output of the
   SMBH. Current X-ray observations from
   \emph{Chandra} have found evidence that AGN feedback exists in the
   local universe \cite{fabian03}. However, X-ray calorimeters on future X-ray
   satellites will be needed for further improvement as they provide the high throughput and spectral
   resolution needed to measure column densities and velocities of
   ionized gas, and consequently the kinetic energy flux, in a large sample of AGN across a broad redshift range.  
  \item
    {\bf Improvements in resolution and sub-resolution modeling for
      direct hydrodynamic simulations.}
    Full hydrodynamic cosmological simulations offer the most promising
    avenue for providing a physically motivated BHMF without free
    parameters, and for unambiguously identifying the relevant
    physical processes in building up the BHMF. However, they
    currently cannot resolve scales relevant to the accretion flow
    onto the SMBH. Numerical codes based on \emph{Adaptive Mesh
      Refinement} techniques will provide improvement in resolution, but it will
    likely be a while before hydrodynamic cosmological simulations are
    able to follow SMBH growth in large cosmological volumes while
    simultaneously resolving the scales relevant for individual black
    holes. In 
    the meantime, further improvement can be made to the
    sub-resolution modeling employed by current hydrodynamic
    simulations. 

    One way of improving current sub-resolution models may be to
    implement the results on AGN feedback based on the type of work
    described in the previous bullet-point. Another improvement is in
    modifying the sub-resolution model for the SMBH accretion
    flow. Current methods assume the Bondi rate combined with a correction
    factor to account for the fact that the temperature and density of the
    gas are not resolved at the Bondi radius. Not surprisingly, the growth
    of the SMBH is sensitive to how this correction factor is modeled
    \cite{booth09}. Moreover, sub-resolution models based on the Bondi 
    rate neglect the angular momentum of the gas, and thus the Bondi rate
    may not be representative of the actual accretion rate onto the
    SMBH. Hopkins \& Quataert \cite{hopkins11} used high-resolution hydrodynamic simulations
    to conclude that the Bondi rate was a poor estimate of the actual
    accretion rate onto the SMBH, and describe a sub-resolution model
    which accurately estimated the actual accretion rate in their
    simulations. In addition, Power, Nayakshin, \& King \cite{power11} suggest an alternative
    sub-resolution model based on an `accretion particle' to provide a
    more accurate estimate of the black hole accretion rate. The
    implementation of improved sub-resolution models for accretion rate
    and feedback into cosmological hydrodynamic simulations, as well as
    further improvements to the sub-resolution models, will result in more
    accurate predicted BHMFs, allowing a more insightful comparison with
    empirical BHMFs.
  \end{itemize}

\begin{figure}[t]
  \includegraphics[scale=0.25,angle=-90]{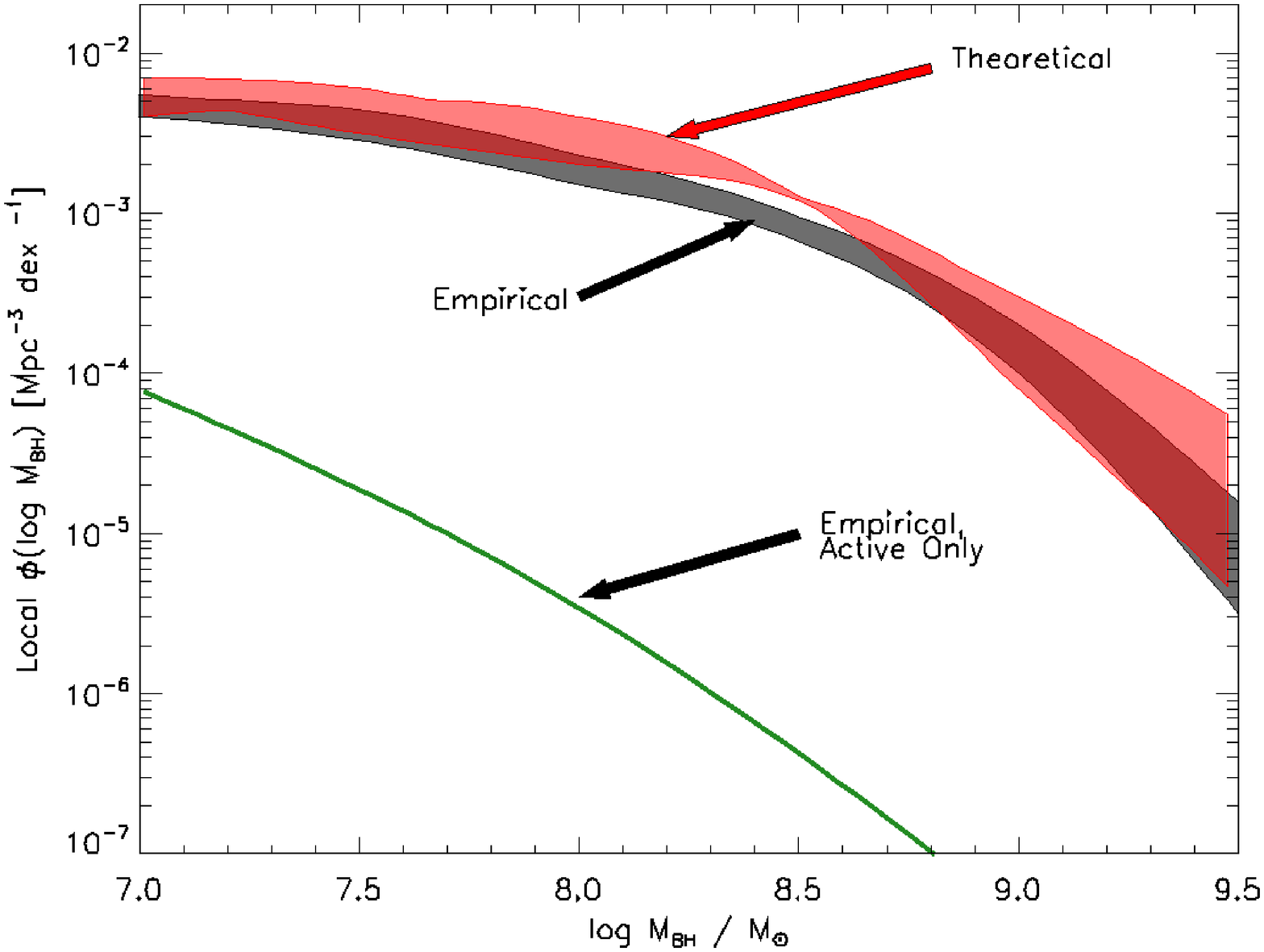}
  \includegraphics[scale=0.25,angle=-90]{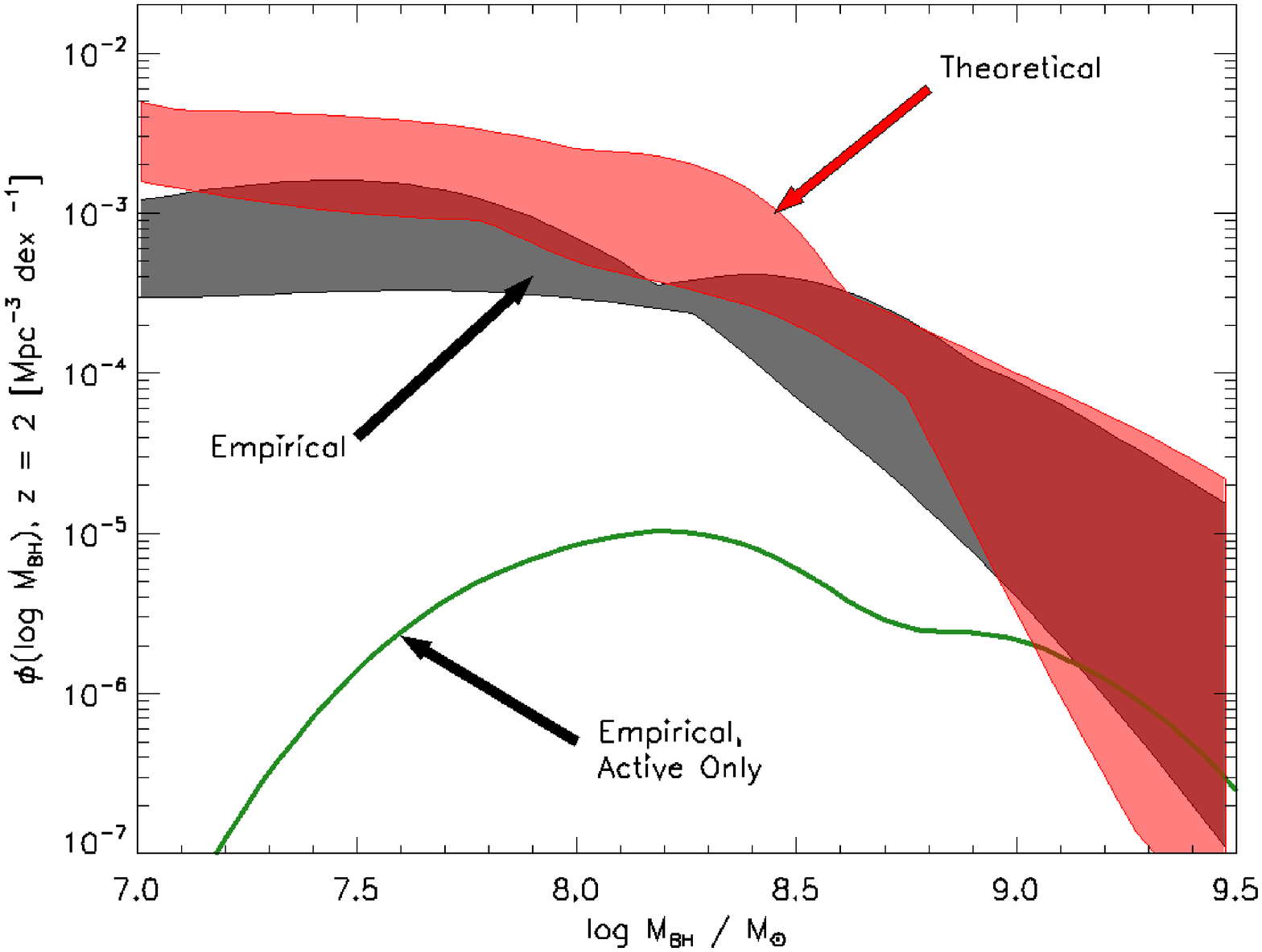}
  \caption{Comparison of empirical estimates of the BHMF (grey shaded
    region) with BHMFs predicted by theoretical models for SMBH
    formation and growth (red shaded region), in both the local
    universe (left) and at $z = 2$ (right). Also shown are the
    estimated BHMF of broad line AGN only (solid green line), from
    Schulze \& Wisotzki \cite{schulze10} (left) and Kelly et
    al. \cite{kelly10} (right). The empirical estimates of the BHMF
    are those shown in Figures \ref{f-lcurve} and \ref{f-mh08_bhmf},
    while the theoretical estimates are those shown in Figure
    \ref{f-theory}. The shaded regions define the spread in the
    estimates and models, and may be considered to be a crude estimate
    of their uncertainty. In general, the theoretical number densities
    are consistent with the empirical ones to within a factor of a
    few.}
  \label{f-compare}
\end{figure}

  In this paper we have reviewed current estimates of the SMBH mass
  function, as well as theoretical models for the BHMF. As discussed
  above, each of the methods for estimating the BHMF has their own set
  of systematics. In Figure \ref{f-compare} we compare the empirical
  estimates of the local BHMF (defined by the shaded region in Figure
  \ref{f-local_bhmf}) with the BHMFs predicted by the theoretical
  models compiled in Figure \ref{f-theory}. In addition, in Figure
  \ref{f-compare} we also compare the empirical BHMFs at $z = 2$, as
  estimated using the lightcurve method (shown in Figure
  \ref{f-lcurve}) and the black hole fundamental plane (shown in
  Figure \ref{f-mh08_bhmf}), with BHMFs predicted by the theoretical
  models. In both Figures we also include the BHMFs for broad line AGN
  as estimated by Schulze \& Wisotzki \cite{schulze10} and Kelly et
  al. \cite{kelly10}. In general, the theoretical models are
  consistent to within a factor of a few with the empirical estimates
  of the BHMF, although there is a large spread in the models and
  empirical estimates at $z = 2$. Moreover, the estimated number
  densities of broad line AGN are significantly lower than those of
  all SMBHs, suggesting that only a small fraction of SMBHs are active
  across a broad range in $M_{BH}$, except for possibly SMBHs at $z
  \sim 2$ with $M_{BH} \gtrsim 10^9 M_{\odot}$.

Despite the differences in the methods for estimating the BHMF, and
the theoretical models, they have lead to a number of common
conclusions. In particular, the empirical results have presented a
picture whereby SMBHs grow primarily via accretion in active phases
(Eddington ratios $L_{bol} / L_{Edd} > 0.01$), that quasar activity is
a relatively short-lived phenomenon relative to the lifetime of the
SMBH and host galaxy (i.e., small `duty cycles' for AGN activity), and
that SMBH growth is anti-hierarchical with the most massive end of the
BHMF being built up first. These empirical results are qualitatively
in agreement with the steadily improving theoretical models.

  We would like to thank Tommaso Treu, Priya Natarajan, Marta
  Volonteri, Aneta Siemiginowska, Xiaohui Fan, and Marianne
  Vestergaard for helpful comments on an earlier draft of this
  review. In addition, we would like to thank two anonymous referees
  whose comments improved our review. BK acknowledges support by NASA through Hubble Fellowship
  grant \#HF-51243.01 awarded by the Space Telescope
  Science Institute, which is operated by the Association of
  Universities for Research in Astronomy, Inc., for NASA, under contract
  NAS 5-26555.

\end{document}